\documentclass[12pt, twoside]{article}
\usepackage{a4wide,amssymb,cite}
\usepackage{epsfig,axodraw}

\def\de{\partial}

\def\g{\gamma}

\def\d{\delta}

\def\e{\eta}
\def\la{\lambda}

\def\m{\mu}
\def\n{\nu}
\def\r{\rho}

\def\p{\pi}
\def\f{\phi}

\def\th{\theta}

\newcommand{\be}{\begin{equation}}
\newcommand{\ee}{\end{equation}}
\newcommand{\bea}{\begin{eqnarray}}
\newcommand{\eea}{\end{eqnarray}}
\newcommand{\beqar}{\begin{eqnarray*}}
\newcommand{\eeqar}{\end{eqnarray*}}
\newcommand{\eg}{{\it e.g.,}\ }
\newcommand{\ie}{{\it i.e.,}\ }
\newcommand{\reef}[1]{(\ref{#1})}

\begin{document}

\begin{titlepage}

\rightline{hep-th/0602208}
\rightline{DESY-05-255}
\rightline{February 2006}

\begin{centering}
\vspace{1cm}
{\large {\bf Scalar mode analysis of the warped Salam-Sezgin model }}\\

\vspace{1.5cm}

 {\bf Hyun Min Lee}$^{a,*}$ and {\bf Antonios~Papazoglou} $^{b,**}$ \\
\vspace{.2in}

$^{a}$ Deutsches Elektronen-Synchrotron DESY, \\ Notkestra\ss e
        85, 22607 Hamburg, Germany. \\
\vspace{3mm}
$^{b}$ \'Ecole Polytechnique F\'ed\'erale de Lausanne,\\ Institute of Theoretical Physics,\\
SB ITP LPPC BSP 720, CH 1015, Lausanne, Switzerland.

\end{centering}
\vspace{2cm}

\begin{abstract}

We study the scalar perturbation sector of the general
axisymmetric  warped Salam-Sezgin  model with codimension-2 branes.
We focus on the perturbations which mix with the dilaton.
We show that the scalar
fluctuation analysis can be reduced to studying two scalar modes
of constant wavefunction, plus modes of non-constant
wavefunction which obey a single
Schr\"odinger equation. 
From the obtained explicit solution of the scalar modes, we point out the
importance of the non-constant modes in describing the four dimensional effective
theory. This observation remains true for the unwarped case and was
 neglected in the relevant literature.  Furthermore, we show that the warped solutions are free of instabilities.

\end{abstract}

\vspace{1cm}
\begin{flushleft}

$^{*}~$ e-mail address: hyun.min.lee@desy.de \\
$ ^{**}$ e-mail address: antonios.papazoglou@epfl.ch

\end{flushleft}
\end{titlepage}

\section{Introduction}

Among six dimensional supergravities, the Salam-Sezgin model \cite{SS} (the supersymmetric analogue of \cite{Randjbar-Daemi:1982hi}) has received particular attention for the past decades. It has the attractive feature that it gives rise to a massless  chiral gravitino in four dimensions, thus reducing the supersymmetry to $N=1$. By enlarging the gauge group and adding a number of hypermultiplets \cite{RSS} the model can be rendered anomaly free\footnote{See \cite{newanomalyfree} for recent advances on constructing anomaly free models in six-dimensional supergravities.} via the Green-Schwarz mechanism. The later model is more relevant to phenomenology since it contains a large gauge group with matter fields charged under it. Recently, there has been renewed interest in the Salam-Sezgin model, due to the fact that the vacua that are obtained, exhibit the mechanism of spontaneous compactification with gauge field fluxes.

Another particular characteristic of this model, that was noticed recently \cite{gibbons}, is that  all the non-singular (\ie with no singularities worse than conical) maximally symmetric vacua are of the type (Minkowski)$_4 \times X_2$, with $X_2$ a two dimensional manifold. The generic vacuum solutions of this type
have been found to get a warping in front of the four dimensional line
element \cite{gibbons,burgess03,Burgess}.
This warping leads always to the appearance of conical singularities which have to be supported by codimension-2 branes.
The unique vacuum that preserves $N=1$ supersymmetry is the one 
that has no warping 
(and with no branes present) in the four dimensional world-volume
and that has its gauge field flux embedded in the
gauged $U(1)_R$ direction. The appearance of the warping breaks
the remaining $N=1$ supersymmetry. More general
warped solutions without axial symmetry of extra dimensions have been
found in Ref.~\cite{leelud}.

The importance of such kind of compactifications has been increased
the last years with the consideration of models which try to ameliorate
the cosmological constant problem,
the so called selftuning models \cite{selftune}.
In these models, use was
made of a special property of the codimension-2 branes,
that they do not curve their world-volume, but instead induce a deficit angle
in the bulk \cite{luty}.
Thus, the vacuum energy of fields living on these branes
does not gravitate in four dimensions. This would give the hope to solve the puzzle of the smallness of the cosmological constant.
However, from the available models,
the ones with flux compactifications have a hidden fine-tuning
related to the flux quantization or conservation
condition \cite{fine-tuning,gibbons,Burgess,porrati}.
The cosmology on a thick codimension-2  brane in six dimensional  flux models
has been studied
in Ref.~\cite{cline} and has come to the same conclusion regarding the fine-tuning.
Recently, different kinds of compactifications
with de Sitter (singular) 3-branes \cite{burgessnew} or a 4-brane
only \cite{wiltshire} have been put forward in the Salam-Sezgin model.

In the present paper, we will discuss the linearized scalar fluctuations
of the Salam-Sezgin model for the general warped
background of the form given in Refs.~\cite{gibbons,Burgess}. A similar analysis for a non-supersymmetric model has been recently done in \cite{jap}. 
Although we focus on the warped solution with axial symmetry of extra
dimensions, our fluctuation analysis is also applied
to the general warped background without axial symmetry
in the local patch coordinate for each brane.
We will study only the fluctuations which are coupled with the dilaton
perturbations.
We see that they are divided to fluctuations with constant profile along the extra dimensions and to fluctuations with non-trivial wavefunctions. The lowest massive non-constant mode mixes however with the massive constant mode. This mixing is always present irrespective of the presence of warping and has been neglected in the literature when discussing the effective
four-dimensional physics of the model \cite{Quev,gibbons,porrati}.
In particular, in the four-dimensional supergravity
description of the unwarped solution without branes \cite{Quev},
the new non-constant mode has to be included
as a new massive chiral multiplet relevant for low energy physics.
Moreover, we can see that the solutions of general warping are free of instabilities.

in the warped case, we have found the interesting result 
that the mixing with the new mode plays a crucial role in 
determining the instability of the solution, from the wrong sign of the kinetic term of one mode for some region of the parameter space.

The paper is organized as follows. 
In the beginning we will review the Salam-Sezgin model and its vacua in general (warped and unwarped). In the following we will derive the perturbation equations and reduce them to a single Schr\"odinger. Then, we will present the explicit 
wavefunctions and masses for scalar modes. 
Next, we will derive the effective action, showing the mixing of the 
constant massive mode and lowest massive non-constant mode 
and comment on the stability of the warped solution. 
Finally, the conclusions will be drawn. 
The detailed derivation of the linearized equations and the quadratic effective action 
for the scalar modes is presented in the following two appendices.

\section{Salam-Sezgin model: general axisymmetric vacua}

We will first review the general  vacuum solutions of Salam-Sezgin model \cite{SS} with axial
 symmetry.  The bosonic sector of the system consists of the metric $g_{MN}$, a    Kalb-Ramond field $B_{MN}$, a dilaton $\f$ and a gauge field $A_M$. For the purpose of this paper, we will set the  Kalb-Ramond field to its zero background value,  so we will not include it in the action.  Then, the bosonic bulk action of the system  is given by

\bea
S=\int d^6X\sqrt{-g}\bigg[R-\frac{1}{4}e^{\frac{1}{2}\phi}F^2_{MN}
-\frac{1}{4}(\partial_M\phi)^2-8g^2 e^{-\frac{1}{2}\phi}\bigg]~,
\label{action}
\eea
supplemented with the 3-brane action as

\be
  S_{\rm brane} = -\int d^4x\sqrt{-{\hat g}}\,V_s
 =\int d^6 X\sqrt{-g}\,{\cal L}_4~, \label{brane}
\ee
where a distributional brane energy density is

\begin{equation}
{\cal L}_4=-\int d^4x\sqrt{\frac{-{\hat g}}{-g}}V_s\delta^{(6)}(X-X(x))~.
\end{equation}
Here, $V_s$ is the tension, ${\hat g}_{\mu\nu}$ is the metric pulled back
to the brane worldvolume and $X^M(x)$ is the embedding of the brane in the six-dimensional bulk.
The gauge coupling $g$ corresponds to the gauged $U(1)_R$ of the model and in principle different from the the gauge coupling $\tilde{g}$ of the gauge field $A_M$. The Einstein and field equations derived
from the above action are

\bea
R_{MN}=&&2g^2\,e^{-\frac{1}{2}\phi}g_{MN}
+\frac{1}{2}e^{\frac{1}{2}\phi}(F_{MP}F_N\,^P-\frac{1}{8}g_{MN}F^2_{PQ})
\nonumber \\
&&+\frac{1}{4}\partial_M\phi\partial_N\phi+ {\hat T}^b_{MN}~,
\eea
with ${\hat T}^b_{MN}$ being the brane contribution and

\bea
\square^{(6)}\phi=&&\frac{1}{4}e^{\frac{1}{2}\phi}F^2_{PQ}
-8g^2\,e^{-\frac{1}{2}\phi}~,\label{scalareq}
\\
\partial_M(\sqrt{-g}e^{\frac{1}{2}\phi}F^{MN})=&&0~.\label{gaugeeq}
\eea

Assuming the axial symmetry in the internal space, there will be  in general, two 3-branes sitting in the antipodal points of the axis of symmetry.
The general warped solution in this case can be analytically found in the following gauge \cite{gibbons,Burgess}

\bea
ds^2&=&W^2(r)\eta_{\mu\nu}dx^\mu dx^\nu+\g^2(r)[dr^2+\la^2
\alpha^2(r)d\theta^2]~,\\
\phi&=&4\ln W~,
\eea
with the various functions given by

\bea
\g(r)&=&{W \over f_0}, \ \ \alpha(r)=r ~{f_0 \over f_1}~, \label{m1}\\
F_{mn}&=&\epsilon_{mn}\frac{\la q \g^2 \alpha}{W^6}~, \label{flux}\\
W^4&=&\frac{f_1}{f_0}, \ \ f_0=1+\frac{r^2}{r^2_0}, \label{m2}\ \
f_1=1+\frac{r^2}{r^2_1}~,
\eea
where flux $q$ is a constant and the two radii are given by

\be
r^2_0=\frac{1}{2g^2}~, \quad  r^2_1=\frac{8}{q^2}~.\label{paras}
\ee
The quantization condition of the gauge field flux is given by the relation

\be
\frac{4\la\tilde{g}}{q}=n,  \quad  n={\rm integer}~.\label{fluxq}
\ee

For the given metric solution, the brane contribution to the Einstein equation
is expressed by

\begin{equation}
{\hat T}^b_{MN}=-\frac{1}{2}\frac{\sqrt{-{\hat g}}}{\sqrt {-g}}V_s\,
({\hat g}_{\mu\nu}\,\delta^\mu_M\delta^\nu_N-g_{MN})
\delta^{(2)}({\vec y}-{\vec y}_s)~,
\end{equation}
where $y_s$ is the brane position. The definition of the delta function is such that $\int d^2y \d^{(2)}({\vec y})=1$, and thus if expressed in the above coordinates (with $r$ the radial coordinate), it is $\d^{(2)}({\vec y})=\d(r)/(2\p)$.
In this general solution, the metric has two conical singularities,
one at $r=0$ and the other at $r=\infty$\footnote{Note that this point is at a finite proper distance from $r=0$.},
with deficit angles $\d_s$ (supported by tensions $V_s=2 \d_s$) given by

\bea
\frac{\delta_0}{2\pi}&=&1-\la~,\\
\frac{\delta_\infty}{2\pi}&=&1-\la \frac{r^2_1}{r^2_0}=1-{n^2 \over \la}\left(\frac{g}{\tilde{g}}\right)^2~.
\eea

For $r_0=r_1$, \ie for $q=4g$, we have the unwarped model. In the case when $\la=1$ and $\tilde{g}=g$, the unwarped case is possible only if  $n=1$, \ie with no branes present. In all cases where 3-branes are present in the vacuum, supersymmetry is completely broken.

Finally, let us go to a coordinate system which is Gaussian-normal with respect to the two branes. In this  new  radial coordinate the perturbation equations in the next section will be  expressed in a more convenient way. Thus, if we define

\be
d\r = \g dr , \ \ a= \g \alpha~,\label{m3}
\ee
the metric is expressed as

\be
ds_6^2=W^2 \e_{\m\n}dx^\m dx^\n + d\r^2+\la^2 a^2 d\th^2~.
\ee

The background equations of motion are expressed in the new coordinate $\r$ as (all primes from now on are derivatives with respect to $\r$)

\bea
{W' a' \over Wa}&=&{W'' \over W}+{W'^2 \over W^2}~,\\
{W'^2 \over W^2}+{W' a' \over Wa}&=&{1 \over W^2}\left[-g^2+{q^2 \over 16 W^8} \right]~,\\
{a'' \over a}+4 {W' a' \over Wa}&=&-{1 \over W^2}\left[2g^2+{3q^2 \over 8 W^8}\right]-{V_s \over 4\p \la a}\d(\r- \r_s)~,
\eea
where we have substituted

\be
\f=4 \ln W ~~~~ {\rm and} ~~~~ F_{\r\th}={\la q a \over W^6}~.
\ee
In the unwarped case we can make this change of coordinate 
explicitly and have that

\be
a(\r)={r_0 \over 2}\sin \left(2 \r \over r_0\right)~.
\ee

\section{Linearized scalar perturbations}

We would like in this section to perturb the above general vacuum solution and in particular the spin-0 sector. For this purpose we consider the following ansatz for the perturbed metric:
\be
ds_6^2=e^{-\psi}W^2 \e_{\m\n}dx^\m dx^\n + e^{\xi} (d\r^2+e^{2(\psi-\xi)}\la^2 a^2 d\th^2)~. \label{metrpert}
\ee

The perturbation in front of $d\th^2$ is the right one to avoid mixing with the graviton - or in other words it comes from the $(\m\n)$ Einstein equations with $\m \neq \n$ (see \eg \cite{ClineGiova}). The gauge field perturbation is considered  as following
\be
F_{\m \th}=\nabla_\m a_\th~~~~,~~~~F_{\r\th}={\la q a \over W^6}+a_\th'~, \label{gaugepert}
\ee
with all the other components vanishing. In addition, the scalar field perturbation is

\be
\f=4 \ln W +f~. \label{phipert}
\ee

In all the above perturbations $\psi$, $\xi$, $f$, $a_\th$ are functions of the 4d coordinates ($x$) and the radial one ($\r$). We will  not consider the $\th$-dependence which  will provide the angular excitations of the resulting modes, \ie we will restrict ourselves to the $s$-mode of the excitations.

The above scalar perturbations are a subset of the most general perturbations. The most general perturbations would include also $A_\r$, $A$ (with $A_\m=\de_\m A$), $\zeta \equiv g_{\r\th}$, $b \equiv B_{\r \th}$ and $B$ (the scalar dual of $B_{\m\n}$)  in addition. However, as it was shown in \cite{giovannini}, the complete set of perturbations is divided into two subsets of coupled perturbations with no dynamical mixing between the two subsets\footnote{The proof in \cite{giovannini} does not include the scalar modes, $b$ and $B$, of the Kalb-Ramond field $B_{MN}$. However, it can be easily shown that for angle-independent perturbations,
they have no mixing with  the subset of $\psi$, $\xi$, $f$ and $a_\th$.}. These are the following

\be
\{\psi,\xi, f, a_\th\}~~~~{\rm and}~~~~\{A_\r,A,\zeta,b,B\}~.
\ee
Thus, we make no mistake ignoring the latter subset in this paper to study exclusively the first one. Furthermore, it was shown in \cite{giovannini} that the  perturbations in the above gauge (which is called in \cite{giovannini}, longitudinal gauge) coincide with the gauge invariant perturbations. Thus, there will be no gauge ambiguities in our results.

\subsection{Linearized equations of motion}

 The linearized Ricci and energy-momentum tensors resulting from the above metric and field  perturbations are given in Appendix A. 
Using them, we can write down the various components of the linearized 
Einstein equations.  The $(\m\n)$ component reads

\bea
&&{\square \psi \over 2W^2}+{1 \over 2}\psi'' +3 {W' \over W}\psi'
+ {W' \over W}\xi'  + {1 \over 2}{a' \over a}\psi'
+ {q \over 4\lambda a W^4}a_\th' \nonumber \\
&&-{1 \over W^2}\left[2g^2 -{q^2 \over 8 W^8}\right]\xi - {q^2 \over 4 W^{10}}\psi +{1 \over W^2} \left[g^2 +{q^2 \over 16 W^8}\right]f=0~.
\label{mneq}
 \eea
The $(\m \rho)$ component reads

\be
\nabla_\m \left[ \psi' +\xi' +2{W'\over W}(\psi+\xi)- 2{a'\over a}(\psi-\xi)-2{W'\over W}f - {q \over \la a W^4}a_\th \right]=0~.
\label{mreq}
\ee
The $(\rho \rho)$ component reads

\bea
&&-{\square \xi \over 2W^2}+\psi''+{1 \over 2}\xi'' +4 {W' \over W}\psi'
+2 {W' \over W}\xi' -2{a' \over a}\psi'  + {3 \over 2}{a' \over a}\xi' -2{W' \over W}f' - {3q \over 4\la a W^4}a_\th' \nonumber \\
&&+{V_s \over 4 \p \la a}\d(\r-\r_s) (\psi-\xi) -{1 \over W^2}\left[2g^2 +{3q^2 \over 8 W^8}\right]\xi + {3q^2 \over 4 W^{10}}\psi -{1 \over W^2} \left[-g^2 +{3q^2 \over 16 W^8}\right]f=0~.~~~~~~~~~
\label{rreq}
\eea
The $(\th \th)$ component reads

\bea
&&{\square \xi \over 2W^2}-{\square \psi \over W^2}-\psi''+{1 \over 2}\xi'' -4 {W' \over W}\psi'
+2 {W' \over W}\xi' + {3 \over 2}{a' \over a}\xi'  - {3q \over 4\la a W^4}a_\th' \nonumber \\
&&+{V_s \over 4 \p\la a}\d(\r-\r_s) (\psi-\xi) -{1 \over W^2}\left[2g^2 +{3q^2 \over 8 W^8}\right]\xi + {3q^2 \over 4 W^{10}}\psi -{1 \over W^2} \left[-g^2 +{3q^2 \over 16 W^8}\right]f=0~.~~~~~~~~~~
\label{ththeq}
\eea
In addition, linearizing the scalar equation \reef{scalareq} we obtain

\bea
&&{\square f \over W^2}+f''+4 {W' \over W}f'+ {a' \over a}f'-4{W' \over W}(\psi' + \xi') \nonumber \\
&&+ {1 \over W^2}\left[8g^2 -{q^2 \over 2 W^8}\right]\xi + {q^2 \over W^{10}}\psi -  {1 \over W^2}\left[4g^2 +{q^2 \over 4 W^8}\right]f -{q \over \la a W^4}a_\th'=0~,
\label{feq}
\eea
while from the gauge field equation \reef{gaugeeq} we get

\be
{\square a_\th \over W^2}+a_\th''+6 {W' \over W}a_\th'-{a' \over a}a_\th'
+{\lambda q a \over W^6}\left[-3\psi'+{f' \over 2}\right]=0~.
\label{atheq}
\ee

From  \reef{mreq} we can solve for $a_\th$ as a function of the other three perturbations
\be
a_\th = {\la a W^4 \over q} \left[ \psi' +\xi' +2{W'\over W}(\psi+\xi)- 2{a'\over a}(\psi-\xi)-2{W'\over W}f \right]~.
\label{ath}
\ee

Since \reef{atheq} does not have any singular ($\d$-function) term, we should have at the boundaries that $a'_\th(\r_s)=0$. Then, from  \reef{ath} we find that there exist two possibilities:

(i) If $\psi=\xi$, then the only condition is that the derivatives of the wavefunctions should vanish at the poles, \eg $\psi'(\r_s)=0$.

(ii) If $\psi \neq \xi$, then we find the stronger condition that the wavefunctions themselves should vanish at the boundary with a limit for \eg $\psi$ to be

\be
\lim_{\r \to \r_s} {\psi(\r) \over a(\r)}=0 ~.
\ee
From the above, we see that since $a(\r) \propto (\r-\r_s) + \dots $, the wavefunction in case (ii) should have an expansion around the singularities as $\psi = C (\r-\r_s)^2 + \dots$, which also implies that $\psi'(\r_s)=0$.

Two of the remaining five equations are trivial. The first trivial combination is $(\rho\rho)+(\th \th)+2(\m\n)$.
The  second trivial equation is the gauge equation. To  show the latter,  we substitute $a_\th$ from (\ref{ath}) to \reef{atheq} and in the resulting expression we substitute $\square f$ from (\ref{feq}), and $\square \psi'$, $\square \xi'$ from the derivatives of (\ref{mneq}) and (\ref{rreq}) respectively. When we do that, we find that some local terms ($\d$-function terms) remain, namely

\be
[W W' (f+\xi-3\psi)-\psi']\d(\r-\r_s)
\ee
which give zero contribution since  $W'(\r_s)=0$ and $\psi'(\r_s)=0$.

Thus, finally we have three remaining equations for the three variables $\psi$, $\xi$, $f$. So, after we have substituted $a_\th$ from (\ref{ath}) and simplified the expressions by means of the background equations of motion, we can rewrite
Eqns.~(\ref{mneq}), (\ref{rreq}), (\ref{feq}) as

\bea
&&{\square \psi \over W^2}+{3 \over 2}\psi''+{1 \over 2}\xi'' +9 {W' \over W}\psi'
+ 5 {W' \over W}\xi'  + {1 \over 2}{a' \over a}\psi'+{3 \over 2}{a' \over a}\xi' -{W' \over W}f' \nonumber \\
&&~~~~~~~~~~~~~~~~~~~~~~~~~~~~~~~~~~~~~~~~~-{4g^2 \over W^2}(2\xi-f) +{V_s \over 4 \p \la a}\d(\r-\r_s) (\psi-\xi)=0 ~,~~~~~~~~ \label{scalar1s}\\
&&{\square \xi \over W^2}-{1 \over 2}\psi''+{1 \over 2}\xi'' + {W' \over W}\psi'
+ 5{W' \over W}\xi'  + {5 \over 2}{a' \over a}\psi'+{3 \over 2}{a' \over a}\xi' +{W' \over W}f' \nonumber \\
&&~~~~~~~~~~~~~~~~~~~~~~~~~~~~~~~~~~~~~~~~~-{4g^2 \over W^2}(2\xi-f) +{V_s \over 4 \p \la a}\d(\r-\r_s) (\psi-\xi)=0 ~,~~~~~~~~ \label{scalar2s}\\
&&{\square f \over W^2}+f''-\psi''-\xi'' +6{W' \over W}f' -10 {W' \over W}(\psi'+\xi')+{a' \over a}f'+{a' \over a}\psi'-3{a' \over a}\xi'  \nonumber \\
&&~~~~~~~~~~~~~~~~~~~~~~~~~~~~~~~~~~~~~~~~~+{8g^2 \over W^2}(2\xi-f) -{V_s \over 2\p \la a}\d(\r-\r_s) (\psi-\xi)=0 ~.~~~~~~~~ \label{scalar3s}
 \eea
[The $\d$-function terms in the above equations can be dropped for both cases of boundary conditions (i) and (ii).]

In order to find the spectrum  of the above system, we have to determine the
 relation between the perturbations which give rise to a single fluctuation equation. To do this, we can rewrite the system as

\be
\hat{{\mathcal O}}X=m^2 X~, \label{Osys}
\ee
with $X= (\psi,\xi,f)$ and $\square X = m^2 X$. Although the operator $\hat{{\mathcal O}}$ does not look Hermitian in this $X$ basis,
we can consider the eigenvalue problem by looking for the linear relations
between the perturbations which are consistent with the linearized
equations. Substituting the following ans\"atze

\be
\xi(x,y)=A(y) \psi(x,y) \ \ \  {\rm and}  \ \ \ f(x,y)= B(y) \psi(x,y)~,
\ee
to the above equations, we have to find the coefficients $A$ and $B$ that consistently collapse the system to a single differential equation. In this procedure, we find two distinct cases, in which both $A$, $B$ are found to be constant.

\vspace{5mm}

{\bf Constant wavefunctions}

\vspace{2mm}

The first case is the one where the wavefunctions  are constant, \ie $\psi=\psi(x)$. Then, it is easy to see from the above system, that there are two possible solutions for $(A,B)$

\be
(A,B)=(1,2) \ \  \  {\rm and} \ \ \ (A,B)=(1,-2)~. \label{constantmodes}
\ee
Thus, we have $\xi=\psi$ (therefore the singularity condition (i) is satisfied trivially) and in addition two possibilities for  $f = \pm 2 \psi$. The first mode corresponds to the massless mode with $\square \psi=0$ and the second one to a massive mode with $\square \psi = 16 g^2 \psi$. For following use, we will call the first mode $\psi_0$ and the second one $\psi_1$. The above shows that the two modes which one finds in the unwarped case \cite{Quev}, maintain their form even when we introduce a warping.

Let us note here that the massless mode, which corresponds to the breathing mode of the internal space, has the same relative wavefunction as the graviton zero mode. By this we mean that in the four dimensional part of the  metric we have

\be
ds_4^2=\{W^2(\r)\e_{\m\n}+W^2(\r)[h_{\m\n}(x)-\psi(x)\e_{\m\n}]\}dx^\m dx^\n~.
\ee
This is in contrast with the five dimensional case of \eg the Randall-Sundrum model, where the relative wavefunctions of the radion and the zero mode graviton 
were different \cite{Charmousis}.

\vspace{5mm}

{\bf Non-constant wavefunctions}

\vspace{2mm}

The second case is the one where the  wavefunctions $\psi(x,y)$  have non-trivial profiles. Then, one can see that the only way that all three equations \reef{scalar1s},  \reef{scalar2s}, \reef{scalar3s} collapse to the same second order equation for $\psi$ is when

\be
(A,B)=(-1,-2) ~,
\ee
in other words, when $\xi=-\psi$ and $f=-2\psi$. In this case we obtain the differential equation for the fluctuation $\psi$

\be
{\square \psi  \over  W^2} +\psi''+\left(6 {W' \over W}-{a' \over a} \right)\psi'=0~. \label{otherlevels}
\ee

As discussed before for the case (ii) of boundary conditions, the wavefunctions at the poles of the internal manifold should satisfy $\psi'(\r_s)=\psi(\r_s)=0$ and have an expansion $\psi = C (\r-\r_s)^2 + \dots$.

\vspace{2mm}

Summarizing, the spectrum of the scalar excitations of the model consists of a zero mode, a first excited state with constant wavefunction and a tower of additional excited states with non-constant  wavefunctions. In the unwarped case,
the gauge field perturbation is zero for the zero mode and the massive constant wavefunction state,
but non-trivial for all the other modes. On the other hand,
in the warped case, the gauge field perturbation is nontrivial
for all the massive states.

\subsection{Solutions for the non-constant modes}

To solve Eqn. \reef{otherlevels} for the non-constant modes we will first separate variables as  $\psi(x,\rho)={\tilde\psi}(x)\chi(\rho)$
with $\Box{\tilde\psi}=m^2{\tilde\psi}$. Then Eqn.~(\ref{otherlevels}) becomes

\be
\chi^{\prime\prime}+\bigg(6\frac{W'}{W}-\frac{a'}{a}\bigg)\chi'
+\frac{m^2}{W^2}\chi=0~. \label{eom32}
\ee
Then, let us do the following transformation

\be
d\rho=W dz ~~~~~~, ~~~~~~ \chi=\left({a \over a_0 W^5}\right)^{1/2}\hat{\chi}~,
\label{redef1}
\ee
where we define

\be
a_0\equiv \frac{r_0}{2}\sin\bigg(\frac{2z}{r_0}\bigg)~.
\ee
Let us note here that the $z$ and the $\r$ coordinates coincide for no warping ($W=1$), and then additionally  $a=a_0$ and $\chi=\hat{\chi}$. With the new variables and wavefunctions substituted in  Eqn.~(\ref{otherlevels}), we find that

\be
\ddot{\hat{\chi}} -{\dot{a}_0 \over a_0}\dot{\hat{\chi}}+\left({1 \over 2}{\ddot{a} \over a}
-{3 \over 4}{\dot{a}^2 \over a^2}
-{15 \over 4}{\dot{W}^2 \over W^2} -{1 \over 2}{\ddot{a}_0 \over a_0}
+{3 \over 4}{\dot{a}_0^2 \over a_0^2}\right)\hat{\chi}+m^2 \hat{\chi}=0~,
\label{z1eq}
\ee
where $\dot{\phantom{a}}\equiv d/dz$.
From Eqns.~(\ref{m1}), (\ref{m2}) and (\ref{m3}),
we know the coordinates are related as

\be
r=r_0 \tan \left({z \over r_0}\right).
\ee
Then, using

\be
a=r {W \over f_1} ~~~~~~, ~~~~~~ W=\left({f_1 \over f_0}\right)^{1/4},
\ee
the terms in the parenthesis in Eqn. \reef{z1eq} cancel and we are left with the Schr\"odinger equation

\be \ddot{\hat{\chi}} -{\dot{a}_0 \over a_0}\dot{\hat{\chi}}+m^2
\hat{\chi}=0~. \label{z2eq} \ee As seen from the above equation,
in the $z$-coordinates, the eigenvalues  of the system depend only
on $r_0$ (\ie on $g$) and are {\it independent} of the warping. The
eigenfunctions $\hat{\chi}$ are as well independent of the
warping, however, the wavefunctions $\chi$ in \reef{redef1} depend
on the warping.

\begin{figure}[t]

\epsfig{file=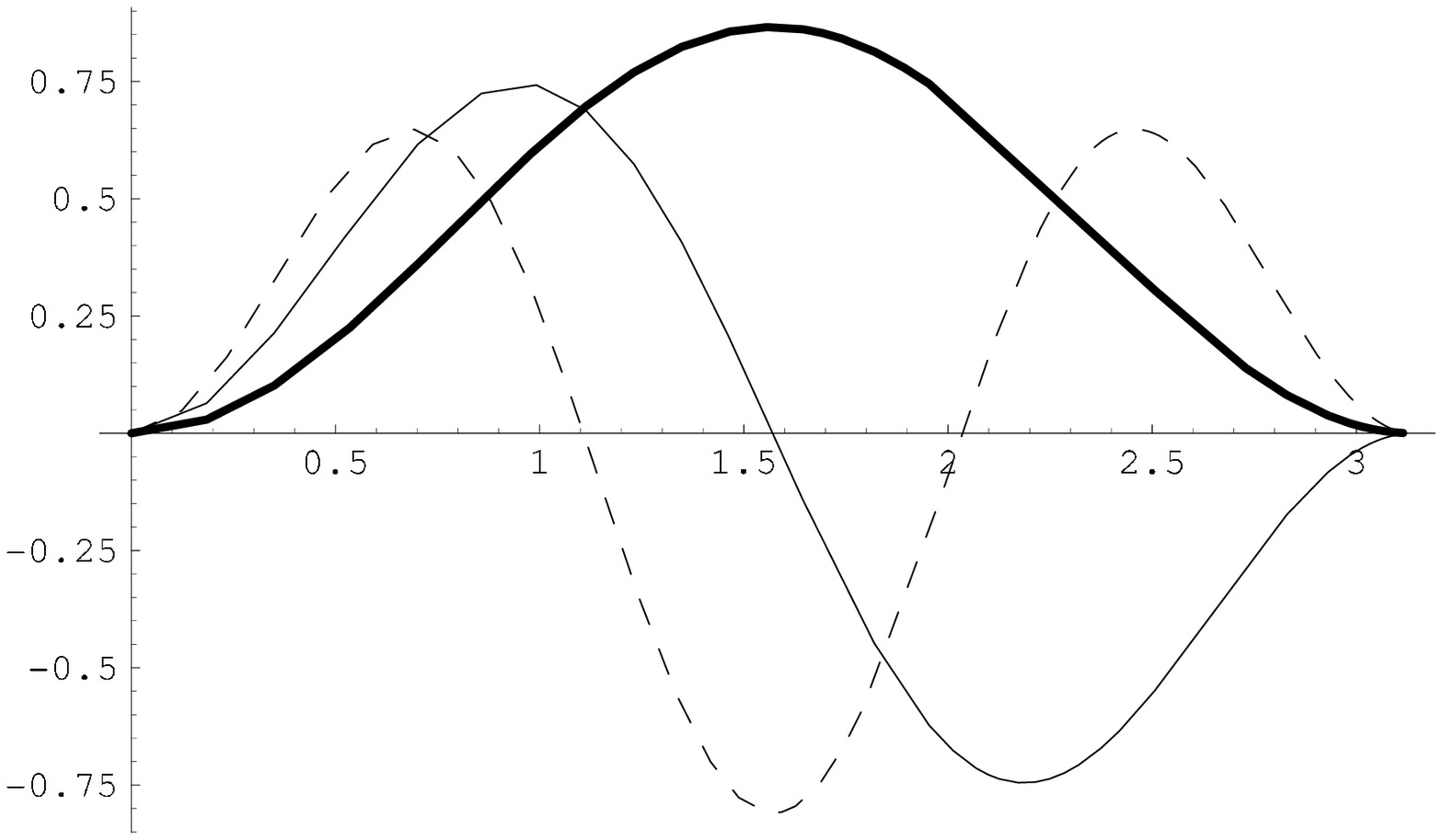,width=5.3cm,height=3.5cm}
\epsfig{file=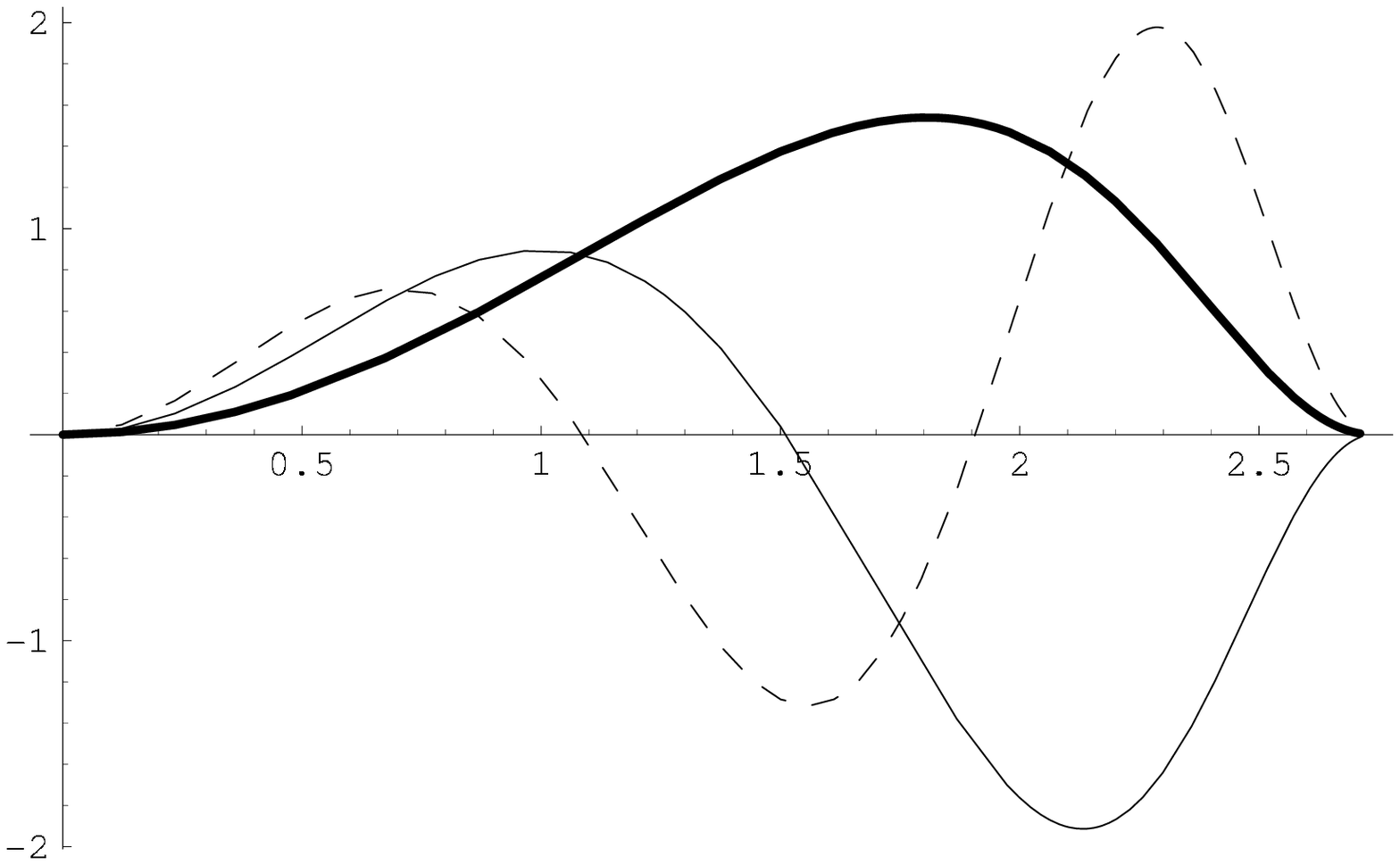,width=5.3cm,height=3.5cm}
\epsfig{file=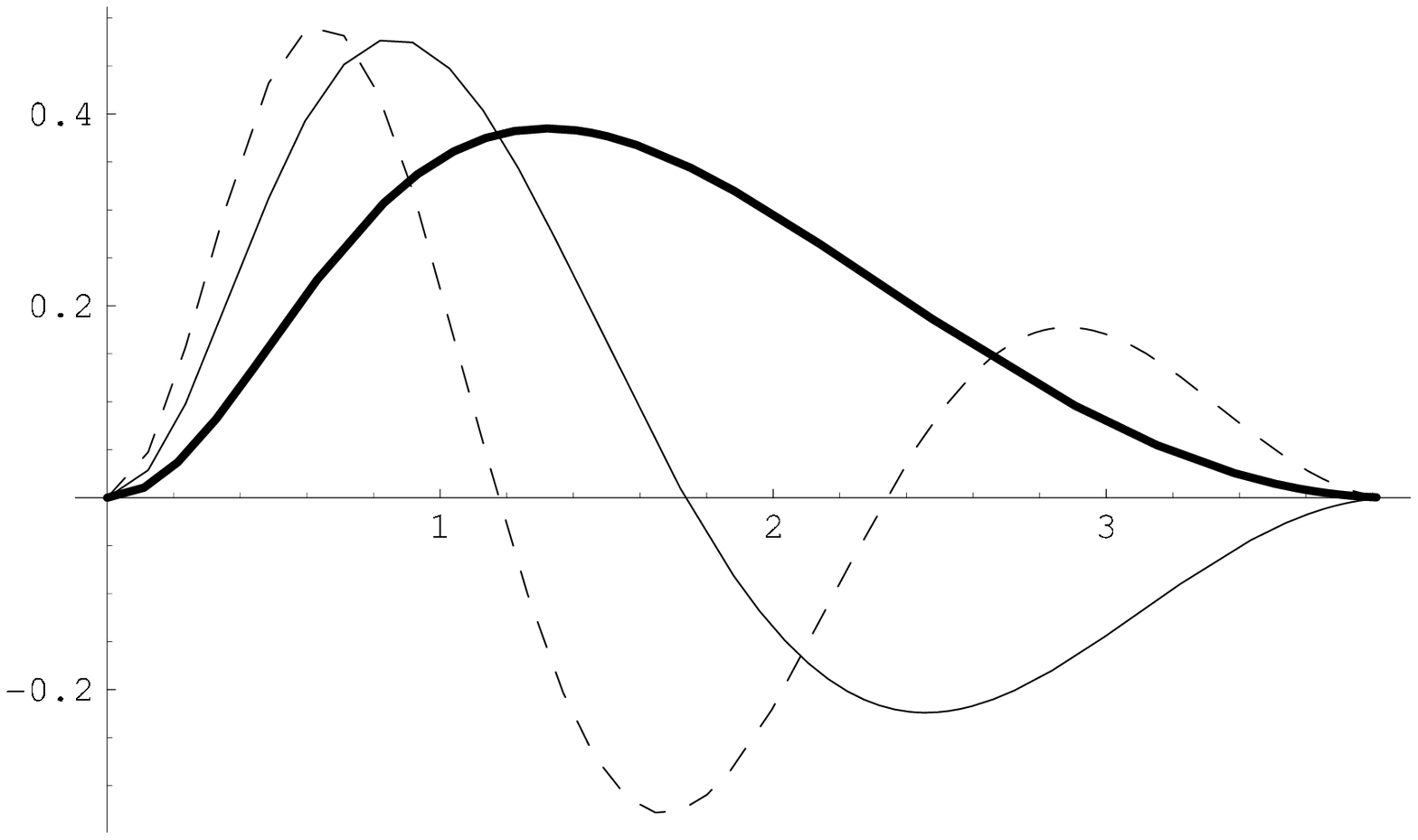,width=5.3cm,height=3.5cm}

\begin{picture}(.01,.01)(0,0)

\Text(80,120)[c]{$4g/q=1$}
\Text(230,120)[c]{$4g/q=2$}
\Text(390,120)[c]{$4g/q=1/2$}

\end{picture}

\caption{The wavefunctions $\chi(\r)$ for first three non-constant modes and  for three different ratios of $4g/q$. The first state  wavefunctions are plotted with thick lines, the second with thin lines and the third with dashed lines.}
\label{3modes}
\end{figure}

The solution of this equation, with the boundary conditions
previously mentioned, is given in terms of the Legendre's
associated polynomials as:

\be
\hat{\chi}_l=\sqrt{\frac{2l-1}{2l(l-1)}}\,(1-y^2)^{1/2} P^1_{l-1}(y) ~~~~~{\rm with}~~~~~y=\cos \left( {2z \over r_0 }\right)\,.
\label{sol}
\ee
On the other hand, the eigenvalues of the equation are given as

\be
m^2_l={4 \over r_0^2}l (l-1) ~~~~~{\rm with}~~~~~l=2,3,4,\cdots
\ee
Here, let us stress that the first level, for $l=2$, has mass $m_2=4g$, which is the same as the one of the constant massive mode $\psi_1$.
Thus, these two degenerate states can {\it mix}, as we will see later. The variable $y$ is related to the original $r$ coordinate as

\be
y=\frac{1-\frac{r^2}{r^2_0}}{1+\frac{r^2}{r^2_0}}~.
\ee
The Legendre's associated polynomials are given as

\be
P^1_n(y)\equiv (1-y^2)^{1/2}\frac{d}{dy}P_n(y)
=\frac{(1-y^2)^{1/2}}{2^nn!}\frac{d^{n+1}}{dy^{n+1}}(y^2-1)^n~,
\ee
satisfying the orthogonality condition 
\bea
\int^1_{-1}dy\, P^1_m(y) P^1_n(y)=\frac{2n(n+1)}{2n+1}\cdot \delta_{mn}~.
\eea

Therefore, we obtain the following orthogonality condition between
scalar modes,
\be
\int {dz \over a_0} \hat{\chi}_m \hat{\chi}_n =\int d\rho\, {W^4 \over a} \chi_m\chi_n =\delta_{mn}~.\label{orthowarp}
\ee

The wavefunctions of the modes with $4g/q>1$ are localised closer to the brane sitting at $r=\infty$, while the   modes with $4g/q<1$ are localised closer to the brane sitting at $r=0$. Examples of these modes are given in Fig.\ref{3modes}.

\section{Effective action for scalar modes}

We would like now to calculate the quadratic four dimensional effective action of the perturbations that we considered in the previous section. In order to do that, we have to expand the  bulk and brane actions (\ref{action}), (\ref{brane}) up to quadratic orders of scalar perturbations, substitute the modes that we have already found and integrate the extra two dimensions. This task is done in detail in Appendix B, leading to the following quadratic effective action for  the scalar modes

\bea
{\cal L}_{\rm eff}&=& 2M^2_P
\bigg[\psi_0\Box \psi_0+A\psi_1(\Box-16g^2)\psi_1 \nonumber \\
&&+B(\psi_1\Box{\tilde\psi}_2+{\tilde\psi}_2\Box\psi_1
-32g^2\psi_1{\tilde\psi}_2)
+C\sum_{l\geq 2}{\tilde\psi}_l(\Box-m^2_l){\tilde\psi}_l\bigg] \nonumber \\
&&+\frac{1}{2}W^2(\rho_s)(\psi_0+\psi_1+\sum_{l\geq 2}{\tilde\psi}_l\chi_l(\rho_s))
\eta^{\mu\nu}T^{\rm brane}_{\mu\nu}~,
\eea
where $M^2_P=\lambda \pi r^2_0$ is the effective four dimensional Planck mass. The constants appearing in the action depend on the ratio  $4g/q$ and read

\bea
A&=&\frac{5}{6}+\frac{1}{12}\bigg(\frac{4g}{q}\bigg)^2
+\frac{1}{12}\bigg(\frac{q}{4g}\bigg)^2~, \\
B&=&\frac{1}{2\sqrt{3}}\bigg[1+\bigg(\frac{4g}{q}\bigg)^2\bigg]~,\\
C&=&\bigg(\frac{4g}{q}\bigg)^2~.
\eea
In the above action, we also added the coupling of the scalar modes to the brane matter
by the brane energy-momentum tensor $T^{\rm brane}_{\mu\nu}$.

Since $\psi_1$ and ${\tilde\psi}_2$ are degenerate in mass, \ie $m^2_2=16g^2$,
we can choose a basis
such that there is no mixing term between these two states in the quadratic action. This is given by 

\be
\phi_\pm=
\frac{1}{\sqrt{2}}\bigg((\mp d+\sqrt{1+d^2})\psi_1\pm {\tilde\psi}_2
\bigg)~, \label{eigen}
\ee
with

\be
d\equiv\frac{C-A}{2B}~.
\ee
Thus,
we obtain the effective action in a diagonal form as

\bea
{\cal L}_{\rm eff}&=&2M^2_P
\bigg[\psi_0\Box \psi_0+
K_+\phi_+(\Box-16g^2)\phi_+
+K_-\phi_-(\Box-16g^2)\phi_- \nonumber \\
&&+\sum_{l\geq 3}{\tilde\psi}_l(\Box-m^2_l){\tilde\psi}_l\bigg] \nonumber \\
&&+\frac{1}{2}W^2(\rho_s)\bigg[\psi_0+\frac{1}{\sqrt{2(1+d^2)}}(\phi_++\phi_-)\bigg]
\eta^{\mu\nu}T^{\rm brane}_{\mu\nu}~,
\eea
with

\be
K_\pm = {\pm B+C(\pm d+\sqrt{1+d^2}) \over \sqrt{1+d^2}}~. \label{Ks}
\ee

These kinetic term coefficients are always positive definite. Therefore the solutions that we described are stable for any value of $4g/q$.

Note also  that there is no coupling of higher KK modes to the brane matter.
So, from the brane perspective, we can think of the four dimensional effective theory, only consisting of the fields $(\psi_0,\phi_+,\phi_-)$. The two massive modes $\f_+$ and $\f_-$, although they are degenerate in mass, they have different couplings to brane fields, since when canonically normalised their couplings will get a contribution from  the different $K_+$ and $K_-$.

In particular,
for the unwarped solution, we have $A=C=1$ and $B=\frac{1}{\sqrt{3}}$.
Then, the effective action for the unwarped case is

\bea
{\cal L}_{\rm eff}&=&2M^2_P
\bigg[\psi_0\Box \psi_0+
(1+\frac{1}{\sqrt{3}})\phi_+(\Box-16g^2)\phi_+
+(1-\frac{1}{\sqrt{3}})\phi_-(\Box-16g^2)\phi_- \nonumber \\
&&+\sum_{l\geq 3}{\tilde\psi}_l(\Box-m^2_l){\tilde\psi}_l\bigg] \nonumber \\
&&+\frac{1}{2}\bigg[\psi_0+\frac{1}{\sqrt{2}}(\phi_++\phi_-)\bigg]
\eta^{\mu\nu}T^{\rm brane}_{\mu\nu}~,
\eea
and the kinetic terms for the scalar modes are positive definite.

\section{Conclusions}

We have analyzed the linearized scalar perturbations (which
couple to the dilaton) of the general axisymmetric vacuum of the
Salam-Sezgin model. We have found that the mass eigenstates
consist of a zero mode with constant wavefunction, a degenerate
pair of first excited states, one with constant and the
other with non-constant wavefunctions, and a tower of heavier
states with non-constant wavefunctions. The degenerate pair has
quadratic mixing in the effective action, even in the case with no
warping. The orthogonal combinations couple to the branes at the
two poles with different strengths. It is important that their kinetic 
terms are positive definite for any value of $4g/q$ and thus the model is stable. These results are expected to hold also for the anomaly free-model \cite{RSS} as long as the additional scalars in the added hypermultiplets are much heavier.

The above-mentioned extra degenerate mode that we have found has to
be included in the effective four dimensional $N=1$ supergravity
description of the unwarped solution without branes \cite{Quev}.
The new chiral multiplet which has to be included will have a
complex scalar component whose real part will be the extra degenerate mode.
The imaginary part will be provided by the scalar perturbation
sector that we did not study in this paper, most probably from the
Kalb-Ramond field. A re-analysis of the effective theory is
necessary to capture the influence of the new mode in the low
energy physics.

Furthermore, it would be interesting to study again the
cosmological properties of this model, as in \cite{Maeda},
including the non-constant mode. The analysis will be more
complicated since the new mode dependence on $\r$ will complicate
the derivation of a four dimensional potential. We plan to address
this issue in a future work.

\section*{Acknowledgments} 
 
We would like to thank C. Charmousis, L. Covi and C. Ludeling for useful discussions. We would also like to thank Y. Burnier for his advice when initially trying to solve numerically the system of Eqns. \reef{scalar1s}, \reef{scalar2s}, \reef{scalar3s}.

\section*{Note added}

In the original (and published \cite{published}) version of the paper, there has been a typo in \reef{eigen} and \reef{Ks}, which indicated an instability of the warped solutions for a certain region of the parameter space. This typo  was noted in Appendix B of \cite{Burgess:2006ds}, and was corrected (along with the conclusion related to it) in the present manuscript. The relevant erratum has been as well  published in \cite{err}. Furthermore, it seems that our requirement that the system \reef{Osys} collapses to a single differential equation is too strong and, thus, there
exist modes \cite{Burgess:2006ds} consistent with our boundary conditions
which we have missed. The correct treatment of the system \reef{Osys} reveals
that the spectrum consists of a massless ground state and a tower of
massive modes with triple degeneracy at each level.

\def\theequation{A.\arabic{equation}}
\setcounter{equation}{0}
\vskip0.8cm
\noindent
{\Large \bf Appendix A: Linearized equations for scalar modes}
\vskip0.4cm
\noindent

The linearized Ricci components computed  from  the perturbed field  ans\"atze \reef{metrpert}, \reef{gaugepert}, \reef{phipert}  are

\bea
R_{\m\n}&=&-W^2\left[{W'' \over W}+ 3{W'^2 \over W^2}+ {a'W' \over aW} \right]\e_{\m\n} + W^2\left[{W'' \over W}+ 3{W'^2 \over W^2}+ {a'W' \over aW} \right](\psi+\xi) \e_{\m\n} \nonumber \\
&& + {1 \over 2}\square \psi \e_{\m\n} +{1 \over 2} W^2\left[\psi''+6{W' \over W}\psi'+2{W' \over W}\xi' + {a' \over a}\psi' \right]\e_{\m\n}~, \\
R_{\m \rho}&=&{1 \over 2}\nabla_\m \psi'+{1 \over 2}\nabla_\m \xi' +  {W' \over W} \nabla_\m (\psi+\xi)-{a' \over a} \nabla_\m (\psi-\xi) ~, \\
R_{\rho\rho}&=&-4{W'' \over W}-{a'' \over a}-{1 \over 2W^2}\square \xi +\psi''+{1 \over 2}\xi'' +4{W' \over W}\psi'+2{W' \over W}\xi' -2{a' \over a} \psi'+{3 \over 2} {a' \over a} \xi' ~,~~~~ \\
{R_{\th \th} \over \lambda^2 a^2}&=&-4{a'W' \over aW}-{a'' \over a}-{1 \over W^2}\square \psi +{1 \over 2W^2}\square \xi-\psi''  +{1 \over 2}\xi'' -4{W' \over W}\psi'+ 2{W' \over W}\xi'+{3 \over 2} {a' \over a} \xi'\nonumber \\
&& -2\left({a'' \over a}+4{a' W' \over aW}\right)(\psi-\xi)~.
\eea
where  $\square \equiv \e^{\m\n}\de_\m \de_\n$. The linearized energy-momentum tensor on the other hand is

\bea
T_{\m\n}&=&-\left[ 4W'^2+8g^2+{q^2 \over 2W^8} +{V_s \over 2 \pi \la a}W^2 \d(\r-\r_s) \right]\e_{\m\n}+\left[ 4W'^2+8g^2+{3q^2 \over 2W^8}  \right]\psi \e_{\m\n}  \nonumber \\
&& +4W'^2\xi \e_{\m\n} + \left[ 4g^2-{q^2 \over 4W^8}  \right]f\e_{\m\n} -2WW'f'\e_{\m\n} - {q \over \la a W^2}a_\th' \e_{\m\n} \nonumber \\
&&+{V_s \over \p \la a}W^2 \d(\r-\r_s)\psi\eta_{\mu\nu}~,~~~~~~~  \\
T_{\m \rho}&=& 2{W' \over W} \nabla_\m f + {q \over \la a W^4}\nabla_\m a_\th ~, \\
T_{\rho\rho}&=&{1 \over W^2}\left[4W'^2 -8g^2+{q^2 \over 2W^8} \right]-{1 \over W^2}\left[8g^2-{q^2 \over 2W^8} \right]\xi - {q^2 \over W^{10}}\psi \nonumber \\
&&+{1 \over W^2}\left[4g^2+{q^2 \over 4W^8} \right]f +2{W' \over W}f' +{q \over \la a W^4}a_\th' ~, \\
{T_{\th \th} \over \lambda^2 a^2}&=&{1 \over W^2}\left[-4W'^2 -8g^2+{q^2 \over 2W^8} \right]-{1 \over W^2}\left[8W'^2+16g^2 \right]\psi+{1 \over W^2}\left[8W'^2+8g^2-{q^2 \over 2W^8} \right]\xi \nonumber \\
&&+{1 \over W^2}\left[4g^2+{q^2 \over 4W^8} \right]f -2{W' \over W}f' +{q \over \la a W^4}a_\th' ~.
\eea
The Einstein equations are in our conventions

\be
R_{MN}= {1 \over 2}\left[T_{MN}-{1 \over 4}T g_{MN}\right]~.
\ee

\def\theequation{B.\arabic{equation}}
\setcounter{equation}{0}
\vskip0.8cm
\noindent
{\Large \bf Appendix B: Quadratic action for scalar modes}
\vskip0.4cm
\noindent

By expanding the bulk action (\ref{action}) and the brane action
(\ref{brane}) up to quadratic orders of scalar perturbations, we
get the four dimensional effective Lagrangian as

\bea {\cal L}_{\rm eff}=2\pi \int d\rho\, \lambda aW^4 ({\cal
L}_k+{\cal L}_m +{\cal L}_{\rm int} \label{effpert})~, \eea
 with

 \bea
  {\cal L}_k&=&W^{-2}\bigg[-\frac{1}{2}\xi\Box\psi-\frac{1}{2}\psi\Box\xi
\nonumber \\
&&+\frac{3}{2}\psi\Box\psi+\frac{1}{2}\xi\Box \xi+\frac{1}{4}f\Box f
+\frac{1}{2}\lambda^{-2}a^{-2}W^2 a_\theta\Box a_\theta\bigg]~, \\
{\cal L}_m&=&-\psi'\xi'-\xi'^2-3\psi'^2-(\psi+\xi)\left(2\psi^{\prime\prime}
+\xi^{\prime\prime}+12\frac{W'}{W}\psi'+8{W' \over W}\xi'
+3{a' \over a}\xi'\right) \nonumber \\
&&-{1\over 2 }\lambda^{-2}a^{-2} W^2 a'^2_\theta
-\frac{q}{\lambda a W^4}\bigg(-3\psi+\frac{1}{2}f\bigg)a'_\theta \nonumber \\
&&- {1 \over 4}f'^2+2  {W' \over W} (\psi+\xi) f' \nonumber \\
&&-\left(4 {a'W' \over aW}+4{W'' \over W}+8 {W'^2 \over W^2}
+{a'' \over a}\right)(\psi+ \xi)^2 \nonumber \\
&&-\frac{q^2}{4W^{10}}\bigg(-3\psi+\frac{1}{2}f\bigg)^2
-\frac{4g^2}{W^2}\bigg(\psi+\frac{1}{2}f\bigg)^2
-\frac{V_s}{\pi\lambda a}\psi^2\delta(\rho-\rho_s)~, \\
{\cal L}_{\rm int}&=&\frac{1}{2}W^{-2}\psi\,\eta^{\mu\nu}T^{\rm br}_{\mu\nu}
\frac{\delta(\rho-\rho_s)}{2\pi\lambda a}~. \\
\eea
We can check that the variation of this action reproduces
indeed  the correct linearized equations.

Now, we can write the scalar perturbations as a sum of the modes we
have found in section 3

 \bea
\psi&=&\psi_0(x)+\psi_1(x)+\sum_n\psi_n(x,\rho)~, \\
\xi&=&\psi_0(x)+\psi_1(x)-\sum_n\psi_n(x,\rho)~, \\
f&=&2\psi_0(x)-2\psi_1(x)-2\sum_n\psi_n(x,\rho)~, \\
a_\theta&=&a^{(1)}_\theta+\sum_n a^{(n)}_\theta~,
\eea
with

\be a^{(1)}_\theta=\frac{8\lambda a W^4}{q}\frac{W'}{W}\psi_1 \ \
, \ \ a^{(n)}_\theta=\frac{4\lambda a
W^4}{q}\bigg(\frac{W'}{W}-\frac{a'}{a}\bigg) \psi_n~.
 \ee
  Then,
inserting the above perturbations into Eqn.~(\ref{effpert}), we can
rewrite the 4D effective Lagrangian as

\bea {\cal L}_{\rm eff} &=&2\pi \int d\rho \,\lambda aW^4({\cal
L}_k+{\cal L}_{\rm const} +{\cal L}_{\rm nonconst}+{\cal L}_{\rm
mix}+{\cal L}_{\rm int})~,
\eea
with

 \bea {\cal L}_k&=&2W^{-2}
\bigg[\psi_0\Box
\psi_0+\bigg(1+\frac{16}{q^2}\bigg(\frac{W'}{W}\bigg)^2
W^{10}\bigg)\psi_1\Box\psi_1 \nonumber \\
&&+\bigg(1+ \frac{8}{q^2}
\frac{W'}{W}\bigg(\frac{W'}{W}-\frac{a'}{a}\bigg)W^{10}\bigg)
(\psi_1\Box\psi_n+\psi_n\Box\psi_1) \nonumber \\
&&+2\bigg(1+\frac{2}{q^2}\bigg(\frac{W'}{W}-\frac{a'}{a}\bigg)^2W^{10}\bigg)
\psi_n\Box\psi_l\bigg]~, \label{lagk}\\
{\cal L}_{\rm const}&=&-\frac{W^6}{2q^2}
\bigg(-16g^2+\frac{q^2}{W^8}\bigg)^2\psi^2_1
+\frac{2}{W^2}\bigg(-16g^2+\frac{q^2}{W^8}\bigg)(\psi_0+2\psi_1)\psi_1
\nonumber \\
&&+\frac{4}{W^2}\bigg(6g^2+\frac{q^2}{8W^8}\bigg)(\psi_0+\psi_1)^2
-\frac{q^2}{W^{10}}(\psi_0+2\psi_1)^2-\frac{16g^2}{W^2}\psi^2_0~, \label{lagc}\\
{\cal L}_{\rm nonconst}&=&-4\psi'_n\bigg[1+
\frac{2}{q^2}\bigg(\frac{W'}{W}-\frac{a'}{a}\bigg)^2W^{10}\bigg]\psi'_l
+8\bigg(\frac{W'}{W}-\frac{a'}{a}\bigg)\psi_n\psi'_l
+\frac{2q^2}{W^{10}}\psi_n\psi_l~,~~~~~~ \label{lagnc}\\
{\cal L}_{\rm mix}&=&-2\psi_0\bigg(\psi^{\prime\prime}_n+4\frac{W'}{W}\psi'_n
+\frac{a'}{a}\psi'_n\bigg) \nonumber \\
&&-2\psi_1\bigg[-\frac{32g^2}{q^2}W^8\bigg(\frac{W'}{W}-\frac{a'}{a}\bigg)
\psi'_n-\frac{1}{W^2}\bigg(-16g^2+\frac{q^2}{W^8}\bigg)\psi_n \nonumber \\
&&\quad+\psi^{\prime\prime}_n+2\frac{W'}{W}\psi'_n+3\frac{a'}{a}\psi'_n\bigg]~,
\label{lagmix}\\
{\cal L}_{\rm int}&=&\frac{1}{2}W^{-2}(\psi_0+\psi_1+\psi_n)\,
\eta^{\mu\nu}T^{\rm br}_{\mu\nu}
\frac{\delta(\rho-\rho_s)}{2\pi\lambda a}~.
\eea
[Here we note that
the summation over indices for massive modes is understood.]

Now let us simplify the terms in the action.
First, by using the fact that

\be
\int d\rho\, aW^2\bigg(-16g^2+\frac{q^2}{W^8}\bigg)=0~,
\ee
we obtain that the mass terms for the constant modes reduce to

\bea
{\cal L}_{\rm const}&=&-\frac{32g^2}{W^2}\bigg[1+\frac{W^8}{64g^2q^2}
\bigg(-16g^2+\frac{q^2}{W^8}\bigg)^2\bigg]\psi^2_1 \nonumber \\
&=&-\frac{32g^2}{W^2}\bigg[1+\frac{16}{q^2}\bigg(\frac{W'}{W}\bigg)^2W^{10}\bigg]\psi^2_1~.
\eea
Moreover, using the fact that

\be
\bigg[W^{10}\bigg(\frac{W'}{W}-\frac{a'}{a}\bigg)^2\bigg]'
=W^{10}\bigg(\frac{W'}{W}-\frac{a'}{a}\bigg)\bigg[\frac{q^2}{W^{10}}
+2\bigg(\frac{W'}{W}-\frac{a'}{a}\bigg)^2
+\frac{V_s}{2\pi\lambda a}\delta(\rho-\rho_s)\bigg]~,
\ee
integrating by parts the first and second terms in Eqn.~(\ref{lagnc}) and using the equation of motion \reef{eom32}, we obtain

\bea
{\cal L}_{\rm nonconst}&=&4\psi_n \bigg[1
+\frac{2}{q^2}\bigg(\frac{W'}{W}-\frac{a'}{a}\bigg)^2W^{10}\bigg]
(\psi^{\prime\prime}_l+6\frac{W'}{W}\psi'_l-\frac{a'}{a}\psi'_l) \nonumber \\
&=&-\frac{4m^2_l}{W^2}\bigg[1
+\frac{2}{q^2}\bigg(\frac{W'}{W}-\frac{a'}{a}\bigg)^2W^{10}\bigg]\psi_n\psi_l~.
\eea
The first term in Eqn.~(\ref{lagmix}) vanishes as a surface term.
So, after integrating by parts for the remaining double derivative term
in Eqn.~(\ref{lagmix}), we get

\be
{\cal L}_{\rm mix}=2\psi_1\bigg[\bigg(\frac{32g^2}{q^2}W^8+2\bigg)
\bigg(\frac{W'}{W}-\frac{a'}{a}\bigg)\psi'_n+\frac{1}{W^2}\bigg(-16g^2+\frac{q^2}{W^8}\bigg)\psi_n\bigg]~.
\ee
By integrating the first term by parts,
we finally obtain the mixing mass term as

\be
{\cal L}_{\rm mix}=-\frac{64g^2}{W^2}\bigg[1
+\frac{8}{q^2}\frac{W'}{W}\bigg(\frac{W'}{W}-\frac{a'}{a}\bigg)W^{10}\bigg]
\psi_1\psi_n~.
\ee

Finally, making a separation of variables
as $\psi_n(x,\rho)={\tilde\psi}_n(x)\chi_n(\rho)$,
we can rewrite the effective Lagrangian for scalar modes as

\bea
{\cal L}_{\rm eff}&=& 2M^2_P
\bigg[\psi_0\Box \psi_0+A\psi_1(\Box-16g^2)\psi_1 \nonumber \\
&&+B_n (\psi_1\Box{\tilde\psi}_n+{\tilde\psi}_n\Box\psi_1
-32g^2\psi_1{\tilde\psi_n})
+C_{nl} {\tilde\psi}_n(\Box-m^2_l){\tilde\psi}_l\bigg] \nonumber \\
&&+\frac{1}{2}W^2(\rho_s)(\psi_0+\psi_1+{\tilde\psi}_n\chi_n(\rho_s))\,
\eta^{\mu\nu}T^{\rm br}_{\mu\nu}~, \label{eff4d}
\eea
where $M^2_P=\lambda \pi r^2_0$ and

\bea
A&=&{2\pi \over M^2_P} \int d\rho\, \lambda aW^2\bigg(1+\frac{16}{q^2}\bigg(\frac{W'}{W}\bigg)^2W^{10}\bigg)~, \\
B_n&=&{2\pi \over M^2_P}\int d\rho\, \lambda aW^2\bigg(1+ \frac{8}{q^2}
\frac{W'}{W}\bigg(\frac{W'}{W}-\frac{a'}{a}\bigg)W^{10}\bigg)\chi_n~, \label{Bmix}\\
C_{nl}&=&{4\pi \over M^2_P} \int d\rho\, \lambda aW^2
\bigg(1+\frac{2}{q^2}\bigg(\frac{W'}{W}-\frac{a'}{a}\bigg)^2W^{10}\bigg)
\chi_n\chi_l~.\label{Ccoef}
\eea
It can be seen through some algebra, that the factor multiplying $\chi_n\chi_l$ in \reef{Ccoef} is the proper weight appearing in  \reef{orthowarp}. Similarly, the factor multiplying $\chi_n$ in \reef{Bmix} is the wavefunction of the lowest non-constant mode $\chi_2$ times again the proper weight. Thus, from the obtained solutions for $\chi_l$
in Eqns.(\ref{redef1}) and (\ref{sol}),
the above expressions can be simplified as following

\bea
A&=&{5 \over 6}+{1 \over 12}\left({4g \over q}\right)^2
+{1 \over 12}\left({q \over 4g}\right)^2~, \\
B_n&=&{1 \over 2 \sqrt{3}}
\left[1+ \left({4g \over q}\right)^2\right]\d_{2n}~, \\
C_{nl}&=&\left({4g \over q}\right)^2\d_{nl}~.
\eea

\end{document}